\documentstyle[12pt,epsf]{article}
\newcommand{\tdm}[1]{\mbox{\boldmath $#1$}}

\newcommand{\gev}{{\rm\; GeV}}
\begin{document}
\begin{titlepage}
\pagestyle{empty}
\vspace*{2cm}
\begin{center}
{\large\bf  Probing the QCD pomeron in doubly tagged $e^+ e^-$  collisions }
\vspace{1.1cm}\\

         {\sc J.~Kwieci\'nski}$^a$,
         {\sc L.~Motyka}$^b$
\vspace{0.3cm}\\

$^a${\it Department of Theoretical Physics, \\
H.~Niewodnicza\'nski Institute of Nuclear Physics,
Cracow, Poland}\\
$^b${\it Institute of Physics, Jagellonian University,
Cracow, Poland}
\vspace{0.3cm}\\
\end{center}
\vspace{1.5cm}
\begin{abstract}
We calculate the total cross-section for $\gamma^* \gamma^*$
collisions and for the process $e^+ e^- \rightarrow
e^+ e^-  + hadrons$ with two tagged leptons assuming dominance of the
QCD pomeron exchange.  We
solve the BFKL equation including the dominant subleading
effects generated by the consistency constraint which restricts the
available phase-space of the emitted gluons to the region in
which the virtuality of the exchanged gluons is dominated by
their transverse momentum squared. Estimate of the possible soft
pomeron contribution to the $\gamma^* \gamma^*$ cross-section
is also presented.  At very high CM energies $W$ the calculated total
$\gamma^* \gamma^*$ cross-section exhibits effective power-law
$(W^2)^{\lambda}$  behaviour with $\lambda \sim 0.3$.
We confront our results with the recent
measurements at LEP and give predictions for the energies which
can be accessible at TESLA and at other future linear $e^+ e^-$
colliders.
\end{abstract}
\end{titlepage}
It has been pointed out  \cite{BRODSKY,BARTELSDR,CZYZFL}
that the measurement of the cross section of the doubly tagged  process 
$e^+e^- \rightarrow e^+e^- + hadrons$ , which is controlled by the total
cross-section describing the interaction of two virtual photons,
i.e. the process  $\gamma^*\gamma^* \rightarrow hadrons$  can be a very useful
tool for probing the QCD pomeron (see Fig.~1).  
In the leading logarithmic approximation
the QCD pomeron corresponds to the sum of ladder diagrams with reggeized gluons along
the chain.  This sum is described by the Balitzkij, Fadin, Kuraev, Lipatov
(BFKL) equation \cite{BFKL,GLR}. \\

Important property of the process $\gamma^* \gamma^* \rightarrow hadrons$
is the fact that by a suitable choice of the kinematical
cuts  one can select   the configuration in which $Q_1^2 \sim Q_2^2$
with both $Q_i^2$ being large.  In this kinematical
configuration, when the two relevant scales are comparable the
conventional LO QCD evolution from the scale $Q_1^2$ to $Q_2^2$ is suppressed
and one can expect that the potential increase of the total cross-section
with energy would then be sensitive to the diffusion of transverse momenta
within the gluon ladder which generates the QCD pomeron.
The process $\gamma^* \gamma^* \rightarrow hadrons$ has also the advantage that for large
$Q_i^2$ its cross-section can be entirely calculated perturbatively.  \\

Existing estimates of the total cross-sections of the process $\gamma^*
\gamma^* \rightarrow hadrons$ have been performed within the leading logarithmic
approximation of the QCD pomeron.  It has however been found recently that the
BFKL equation which generates the QCD pomeron
can acquire very  significant non-leading corrections \cite{BFKLNL,RESUM}.
The purpose of this paper is therefore to (re-) analyse the total
$\gamma^* \gamma^*$ cross-section taking the subleading BFKL effects into account.
To be precise we shall base our calculations on the (modiffied) BFKL equation
with the consistency constraint limiting the phase space of the real
gluon emission.  This constraint is based on the requirement that the virtuality
of the exchanged gluons is dominated by their transverse momentum squared.
 Let us remind that the form of the LO BFKL kernel
where the gluon propagators contain only the gluon transverse momentum squared
etc. is only valid within the region of phase space restricted by this
 constraint.  Formally however, the consistency constraint generates
subleading corrections.  It can be shown that at the NLO accuracy it generates
about 70\% of the exact result for the QCD pomeron intercept.  Very important
merit of this constraint is  the fact that it automatically generates
resummation of higher order contributions which stabilizes the solution
\cite{KMSG}. \\

\noindent
\begin{figure}[hbpt]
\begin{center}
\leavevmode
\epsfxsize = 7.5cm
\epsfysize = 5.3cm
\epsfbox[60 545 311 741]{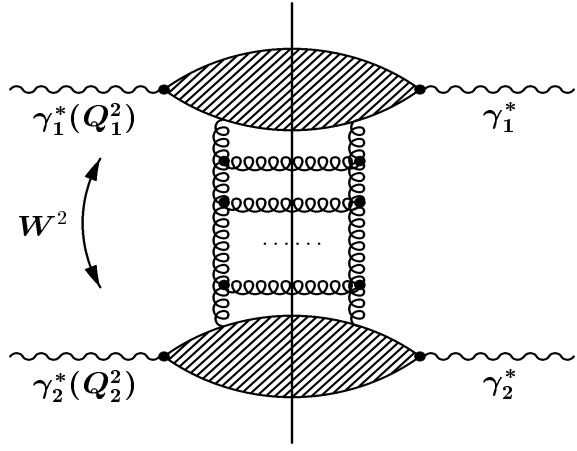}
\end{center}
\caption{\small The QCD pomeron exchange mechanism of the process
   $\gamma_1 ^* (Q_1 ^2) \gamma_2 ^* (Q_2 ^2) \to hadrons$.}
\end{figure}

The cross-section of the process
$e^+e^- \rightarrow
e^+e^- + hadrons$  (averaged over the angle $\phi$ between
the lepton scattering planes in the frame in which the 
virtual photons are aligned along the $z$ axis) is given by the
following formula~\cite{BRODSKY}:
$$
{Q_1^2 Q_2^2 d\sigma \over dy_1 dy_2 dQ_1^2 dQ_2^2} =
\left({\alpha\over 2 \pi}\right)^2
[P^{(T)}_{\gamma/e^+}(y_1)P^{(T)}_{\gamma/e^-}(y_2)
\sigma^{TT}_{\gamma^* \gamma^*}(Q_1^2,
Q_2^2,W^2)+
$$
$$
P^{(T)}_{\gamma/e^+}(y_1)P^{(L)}_{\gamma/e^-}(y_2)
\sigma^{TL}_{\gamma^* \gamma^*}(Q_1^2,Q_2^2,W^2)+
P^{(L)}_{\gamma/e^+}(y_1)P^{(T)}_{\gamma/e^-}(y_2)
\sigma^{LT}_{\gamma^* \gamma^*}(Q_1^2,Q_2^2,W^2)+
$$
\begin{equation}
P^{(L)}_{\gamma/e^+}(y_1)P^{(L)}_{\gamma/e^-}(y_2)
\sigma^{LL}_{\gamma^* \gamma^*}(Q_1^2, Q_2^2,W^2)]
\label{conv}
\end{equation}
where
\begin{equation}
P^{(T)}_{\gamma/e}(y) = {1 + (1-y)^2\over y}
\label{pt}
\end{equation}
\begin{equation}
P^{(L)}_{\gamma/e}(y) = 2{1-y\over y}
\label{pl}
\end{equation}
In Eq.~(\ref{conv}) $y_1$ and $y_2$ are the longitudinal momentum 
fractions of the parent leptons
carried by virtual photons,  $Q_i^2 = -q_i^2$ ($i=1,2$)
where $q_{1,2}$ denote the four momenta of the
virtual photons and $W^2$ is the total CM energy squared of the
two (virtual) photon system, i.e. $W^2=(q_1+q_2)^2$. The cross-sections
$\sigma^{ij}_{\gamma^* \gamma^*}(Q_1^2,
Q_2^2,W^2)$ are the total cross-sections of the process $\gamma^*
\gamma^* \rightarrow hadrons$  and the indices $i,j=T,L$ denote the 
polarization of
the virtual photons.  The functions $P^{(T)}_{\gamma/e}(y)$ and
$P^{(L)}_{\gamma/e}(y)$ are the transverse and longitudinal photon
flux factors.\\

The cross-sections  $\sigma^{ij}_{\gamma^* \gamma^*}(Q_1^2, Q_2^2,W^2)$
are given by the following formula:
$$
\sigma^{ij}_{\gamma^* \gamma^*}(Q_1^2,
Q_2^2,W^2) = P_S(Q_1^2,Q_2^2,W^2)\delta_{iT}\delta_{jT} +
$$
\begin{equation}
{1\over 2 \pi}\sum_q\int_{k_0^2}^{k_{max}^2(Q_2^2,x)} {dk^2\over k^4}
\int _{\xi_{min}(k^2,Q_2^2)}^{1/x} d\xi
 G^{0j}_q(k^2,Q_2^2,\xi)
 \Phi_i(k^2,Q_1^2,x\xi)
\label{csx}
\end{equation}
where
\begin{equation}
k_{max}^2(Q_2^2,x)=-4m_q^2+Q_2^2 \left( {1\over x} - 1 \right)
\label{kmax}
\end{equation}
\begin{equation}
\xi_{min}(k^2,Q^2)=1+{k^2+4m_q^2\over Q^2}
\label{ximin}
\end{equation}
and
\begin{equation}
x={Q_2^2\over 2q_1q_2}
\label{x}
\end{equation}
The functions $G^{0i}_q(k^2,Q^2,\xi)$,
which describe the coupling of the two gluon system to virtual photons
corresponding to the quark box and crossed-box diagrams
are defined as
below \cite{BRODSKY,KMS}
$$
 G^{0T}_q(k^2,Q^2,\xi)=
$$
$$
2 \alpha_{em} \alpha_s(k^2+m_q^2)e_q^2\int_0^{\rho_{max}} d \rho
\int {d^2p^{\prime }\over \pi}\;
\delta\left[\xi-\left(1+{p^{\prime 2}+m_q^2\over z(1-z)Q^2} +
{k^2\over Q^2}\right)\right] \times
$$
\begin{equation}
\left\{ \left[(z^2 + (1-z)^2)\left({\tdm p\over D_1} -
{\tdm p + \tdm k\over D_2} \right)^2 \right]
+m_q^2 \left( {1\over D_1} - {1\over D_2} \right)^2 \right\}
\label{g0t}
\end{equation}
$$
 G^{0L}_q(k^2,Q^2,\xi)=
$$
$$
8 \alpha_{em} \alpha_s(k^2+m_q ^2)e_q^2\int_0^{\rho_{max}} d \rho
\int {d^2p^{\prime }\over \pi}\;
\delta\left[\xi-\left(1+{p^{\prime 2}+m_q^2\over z(1-z)Q^2} +
{k^2\over Q^2}\right)\right]  \times
$$
\begin{equation}
\left[z^2 (1-z)^2 \, \left({1\over D_1} -
{1\over D_2} \right)^2 \right]
\label{g0l}
\end{equation}
where
\begin{equation}
z={1+\rho\over 2}
\label{zlam}
\end{equation}
\begin{equation}
\tdm p=\tdm p^{\prime} + (z-1) \tdm k
\label{pprime}
\end{equation}
$$
D_1=p^2 + z(1-z)Q^2 + m_q^2
$$
\begin{equation}
D_2=(p+k)^2 + z(1-z)Q^2 + m_q^2
\label{d12}
\end{equation}
\begin{equation}
\rho_{max} = \sqrt{1 - {4 m_q ^2 \over {1/x - k^2/Q^2 - 1}}}
\label{rhomax}
\end{equation}
with $m_q$ denoting the quark mass. In our calculations we include
contributions from $u$, $d$, $s$ and $c$ quarks and set $m_u = m_d = m_s = 0$ 
and $m_c = 1.5$~GeV.\\

The function $P_S(Q_1^2,Q_2^2,W^2)$  corresponds to the
the contribution from the region $k^2 \le k_0^2$ in the
corresponding integrals over the gluon transverse momenta.  It
is assumed to be dominated by the soft pomeron contribution
which is estimated from the factorisation of its couplings, i.e.
\begin{equation}
P_S(Q_1^2,Q_2^2,W^2) = {\sigma^{SP}_{\gamma^*(Q_1^2)p}(Q_1^2,W^2)
\sigma^{SP}_{\gamma^*(Q_2^2)p}(Q_2^2,W^2)\over \sigma_{pp}^{SP}(W^2)}
\label{sp}
\end{equation}
We assume that this term is only contributing to the transverse
part. In equation (\ref{sp}) the cross-sections
$\sigma^{SP}_{\gamma^*(Q_i^2)p}(Q_i^2,W^2)$ and
$\sigma_{pp}^{SP}(W^2)$  are the soft pomeron contributions to
the $\gamma^*p$ and $pp$ total cross sections and their
parametrisation is taken from Refs.~\cite{DLTOTCX,DLDIS}.  Their
$W^2$ dependence is, of course, universal i.e.
$$
\sigma_{pp}^{SP}(W^2)=
\beta_p^2\left({W^2\over W_0^2}\right)^{\alpha_{SP}(0)-1}
$$
\begin{equation}
\sigma^{SP}_{\gamma^*(Q^2)p}(Q^2,W^2) = \beta_{\gamma^*}(Q^2)\beta_p
\left({W^2\over W_0^2}\right)^{\alpha_{SP}(0)-1}
\label{sppar}
\end{equation}
with $W_0 = 1\gev$ and $\alpha_{SP}(0) \approx 1.08$.

The function $\beta_{\gamma^*} (Q^2)$ behaves for large $Q^2$ as
$\beta_{\gamma^* } (Q^2) \sim (Q^2)^{-\alpha_{SP}(0)}$ and so the
factorisation formula (\ref{sp}) implies that the soft pomeron contribution
to the total $\gamma^* \gamma^*$ cross-section should behave as:
\begin{equation}
P_S(Q_1 ^2, Q_2 ^2, W^2) \sim
\left( {W^2 W_0 ^2 \over Q_1 ^2 Q_2 ^2} \right) ^{\alpha_{SP}(0)-1}
{1 \over Q_1 ^2 Q_2 ^2}
\label{pssim}
\end{equation}

The functions $\Phi_i(k^2,Q^2,x_g)$ in the second term in Eq.~({\ref{csx})
satisfy the Balitzkij, Fadin, Kuraev, Lipatov  (BFKL) equation   
which, in the leading $\ln(1/x)$ approximation have the following form:
$$
\Phi_i(k^2,Q^2,x_g)=\Phi^0_i(k^2,Q^2,x_g)+\Phi^S (k^2,Q^2,x_g)\delta_{iT}+
{3\alpha_s(k^2)\over \pi} k^2\int_{x_g}^1 {dx^{\prime}\over x^{\prime}}
\int_{k_0^2}^{\infty} {dk^{\prime 2} \over k^{\prime 2}}
$$
\begin{equation}
\left [ {\Phi_i(k^{\prime 2},Q^2,x^{\prime}) - \Phi_i(k^{ 2},Q^2,x^{\prime})
\over |k^{\prime 2} - k^{ 2}|} + {\Phi_i(k^{ 2},Q^2,x^{\prime})\over
\sqrt{4 k^{\prime 4} +  k^{4}}}\right]
\label{bfklll}
\end{equation}
The term proportional to the function $\Phi_i (k'^2, Q^2,x')$ under the
integral corresponds to the real gluon emission while the terms
proportional to $\Phi_i (k^2, Q^2, x')$ to virtual corrections which
are responsible for gluon reggeisation. The inhomogeneous terms
$\Phi_i ^0 (k^2, Q^2, x_g)$ and $\Phi ^S (k^2, Q^2, x_g)$ will be
defined later.\\

In the small $x_g$ limit the solution of the BFKL equation
(\ref{bfklll}) behaves as
\begin{equation}
\Phi_i (k^2, Q^2,x_g) \sim x_g ^{-\lambda_{QCD}}
\label{smallx}
\end{equation}
If in equation (\ref{bfklll}) the running QCD coupling
$\alpha_s (k^2)$ is replaced by the fixed (i.e. $k^2$ independent)
coupling $\tilde \alpha_s$ then the exponent $\lambda_{QCD}$ is given
by the following formula~\cite{BFKL}:
\begin{equation}
\lambda_{QCD} = {6 \tilde\alpha_s \over \pi} [ \psi(1) - \psi(1/2)]
\end{equation}
with $\psi(z) = \Gamma' (z) / \Gamma (z)$, where $\Gamma (z)$ is the
Euler Gamma function and $2[\psi(1)-\psi(1/2)] = 4\ln 2$. The quantity
$1+\lambda_{QCD}$ is the QCD pomeron intercept.\\

In what follows we shall consider the modiffied BFKL equation in
which we restrict the available phase-space in the real gluon
emission by the consistency constraint:
\begin{equation}
k^{\prime 2} \le k^2{ x^{\prime}\over x_g}
\label{cc}
\end{equation}
This constraint follows from the requirement that the virtuality
of the exchanged gluons is dominated by their transverse momentum
squared.
%
%
Modiffied BFKL equations take the following form:
$$
\Phi_i(k^2,Q^2,x_g)=\Phi^0_i(k^2,Q^2,x_g)+\Phi^S(k^2,Q^2,x_g)\delta_{iT}+
{3\alpha_s(k^2)\over \pi} k^2\int_{x_g}^1 {dx^{\prime}\over x^{\prime}}
\int_{k_0^2}^{\infty} {dk^{\prime 2} \over k^{\prime 2}}
$$
\begin{equation}
\left [ {\Phi_i(k^{\prime 2},Q^2,x^{\prime})\Theta
\left(k^2{ x^{\prime}\over x_g}
-k^{\prime 2}\right) - \Phi_i(k^{ 2},Q^2,x^{\prime})
\over |k^{\prime 2} - k^{ 2}|} + {\Phi_i(k^{ 2},Q^2,x^{\prime})\over
\sqrt{4 k^{\prime 4} +  k^{4}}}\right]
\label{bfklcc}
\end{equation}
The consistency constraint (\ref{cc}) lowers the QCD pomeron intercept
since the exponent $\lambda_{QCD}$ can be shown to be now the solution
of the following equation (for the fixed coupling $\tilde\alpha_s$):
\begin{equation}
\lambda_{QCD} = {6 \tilde\alpha_s \over \pi} \left[
\psi(1) - \psi\left({1+\lambda_{QCD} \over 2}\right)\right]
\label{lamcc}
\end{equation}
This equation introduces subleading corrections to the QCD pomeron
intercept and generates their (approximate) resummation to all orders.
In the next-to-leading approximation it exhausts about $3/4$ of the
entire next-to-leading contribution to $\lambda_{QCD}$. 
Another feature of the modiffied
BFKL equation~(\ref{bfklcc}) is the delay of the onset of the
power-law behaviour~(\ref{smallx}) (with $\lambda_{QCD}$ defined now by
Eq.~(\ref{lamcc})) to smaller values of $x_g$ than in the case of the
leading order approximation. This is caused by the fact that for
moderately small values of $x_g$ the (negative) virtual contribution
to the BFKL equation, which is unaffected by the consistency
constraint, dominates over the positive real emission term constrained
by the Theta function in Eq.~(\ref{bfklcc}). 
Both effects (i.e. lowering of the intercept
and delay of the onset of asymptotic small $x_g$ behaviour)
substantially reduce the corresponding cross-section. In what follows
we shall numerically solve equation (\ref{bfklcc}) with the running
coupling constant $\alpha_s(k^2)$.\\

The inhomogeneous terms in equations (\ref{bfklll},\ref{bfklcc})
are the sum of two contributions  $\Phi^0_i(k^2,Q^2,x_g)$
and $\Phi^S(k^2,Q^2,x_g)\delta_{iT}$.  The first contributions
($\Phi^0_i(k^2,Q^2,x_g) $)
correspond to the diagrams in which the two gluon system couples
to virtual photons through a quark box and are  given by
following equations:
\begin{equation}
\Phi^0_i(k^2,Q^2,x_g)=\sum_q \int_{x_g}^1 dz \; \tilde G^0_{iq}(k^2,Q^2,z)
\label{phii0}
\end{equation}
where
$$
\tilde G^0_{Tq}(k^2,Q^2,z)=2\alpha_{em} e_q^2\alpha_s(k^2+m_q^2)
\int_0^1 d\lambda \left\{
{[\lambda^2 + (1-\lambda)^2][z^2+(1-z)^2] k^2 \over \lambda(1-\lambda)k^2+
z(1-z)Q^2 + m_q^2} \right. +
$$
\begin{equation}
2m_q^2 \left. \left[ {1\over z(1-z)Q^2 + m_q^2} -
{1\over \lambda(1-\lambda)k^2+z(1-z)Q^2 + m_q^2}\right] \right\}
\label{tgt0}
\end{equation}
$$
\tilde G^0_{Lq}(k^2,Q^2,z)=16\alpha_{em}Q^2 k^2 e_q^2\alpha_s(k^2+m_q^2)
\times \hfill
$$
\begin{equation}
\int_0^1d\lambda
\left\{ {[\lambda (1-\lambda)][z^2(1-z)^2] \over [\lambda(1-\lambda)k^2+
z(1-z)Q^2 + m_q^2][z(1-z)Q^2 + m_q^2]} \right\}
\label{tgl0}
\end{equation}
The functions $\tilde G^0_{iq}(k^2,Q^2,z)$ are obtained from 
$G^{0i} _q(k^2,Q^2,\xi)$ after integrating over $d\xi$ and unfolding
integration over $d\rho$ in equations (\ref{g0t}) and (\ref{g0l}).
The second term $\Phi^S(k^2,Q^2,x_g)\delta_{iT}$, which is
assumed to contribute only to the transverse component
corresponds to the contribution to the BFKL equation from the
non-perturbative soft region $k^{\prime 2} < k_0^2$.  Adopting
the strong ordering approximation $k^{\prime 2} \ll k^2$ it
is given by the following formula:.
\begin{equation}
\Phi^S (k^2,Q^2,x_g)=
{3 \alpha_s(k^2)\over \pi}
\int_{x_g}^1 {dx^{\prime}\over x^{\prime}}
\int_{0}^{k_0^2} {dk^{\prime 2} \over k^{\prime 2}}
 \Phi_T(k^{\prime 2},Q^2,x^{\prime})
\label{stophi}
\end{equation}
The last integral in equation (\ref{stophi}) can be interpreted
as a gluon distribution in a virtual photon of virtuality $Q^2$
evaluated at the  scale $k_0^2$.   At low values of $x^{\prime}$
it is assumed to be dominated by a soft pomeron contribution and
can be estimated using the factorisation of the soft pomeron
couplings:
\begin{equation}
\int_{0}^{k_0^2} {dk^{\prime 2} \over k^{\prime 2}}
 \Phi_T(k^{\prime 2},Q^2,x^{\prime})=\pi^2 x^{\prime}g_p(x^{\prime},k_0^2)
{\beta_{\gamma^*}(Q^2)\over \beta_p}
\label{inhos}
\end{equation}
where $g_p(x^{\prime},k_0^2)$ is the gluon distribution in a
proton at the scale $k_0^2$
and the couplings  $\beta_{\gamma^*}(Q^2)$ and $\beta_p$
defined by equation (\ref{sppar}). \\

\begin{figure}[h]
\begin{center}
\leavevmode
\noindent {\large a)} \hspace{7.2cm} {\large b)} \hspace{7.2cm}\\ 
\parbox[t]{7.5cm}{
\epsfxsize = 7.5cm
\epsfysize = 7.5cm
\epsfbox[20 255 550 790]{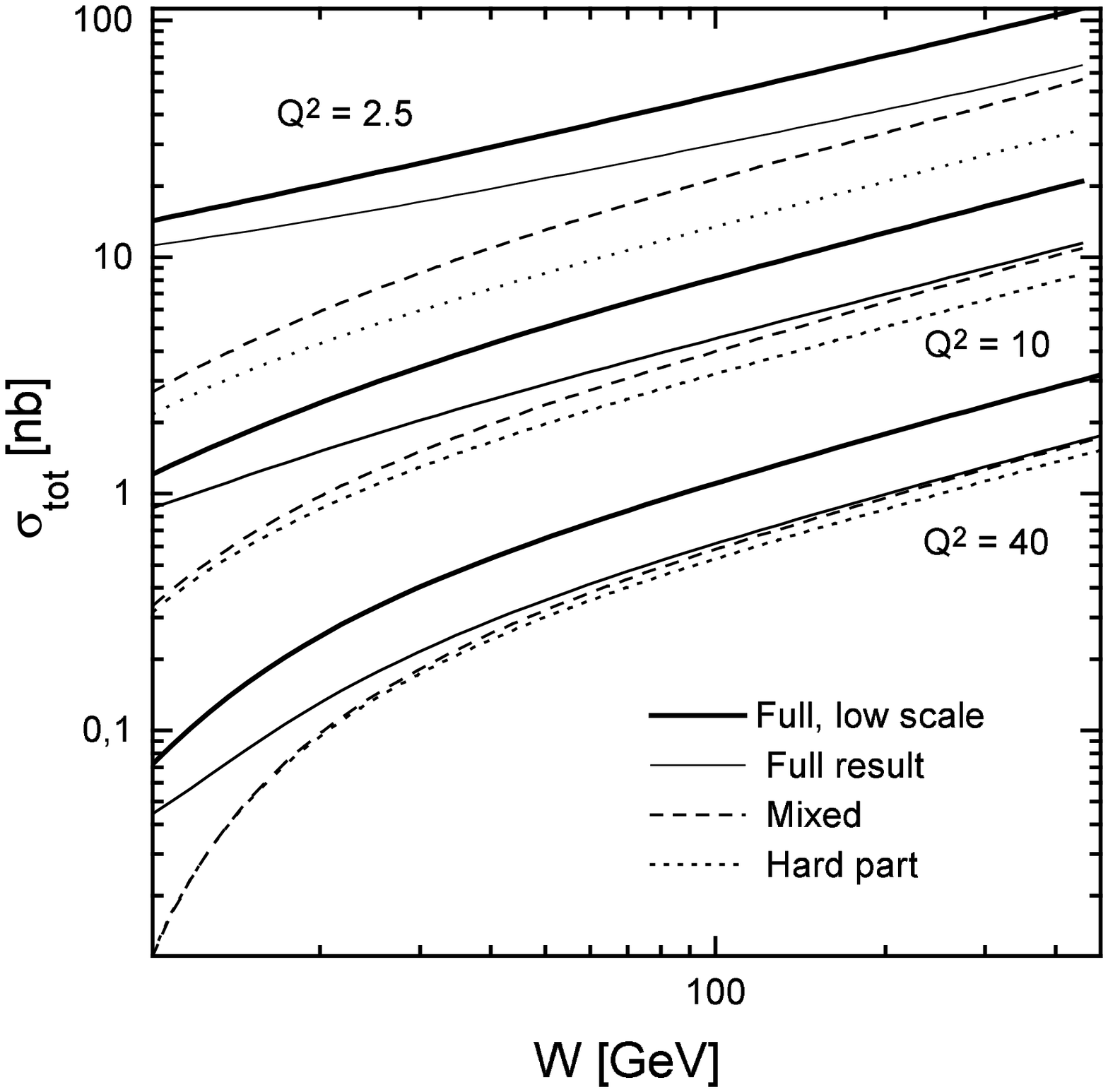}}
\parbox[t]{7.5cm}{
\epsfxsize = 7.5cm
\epsfysize = 7.5cm
\epsfbox[20 255 550 790]{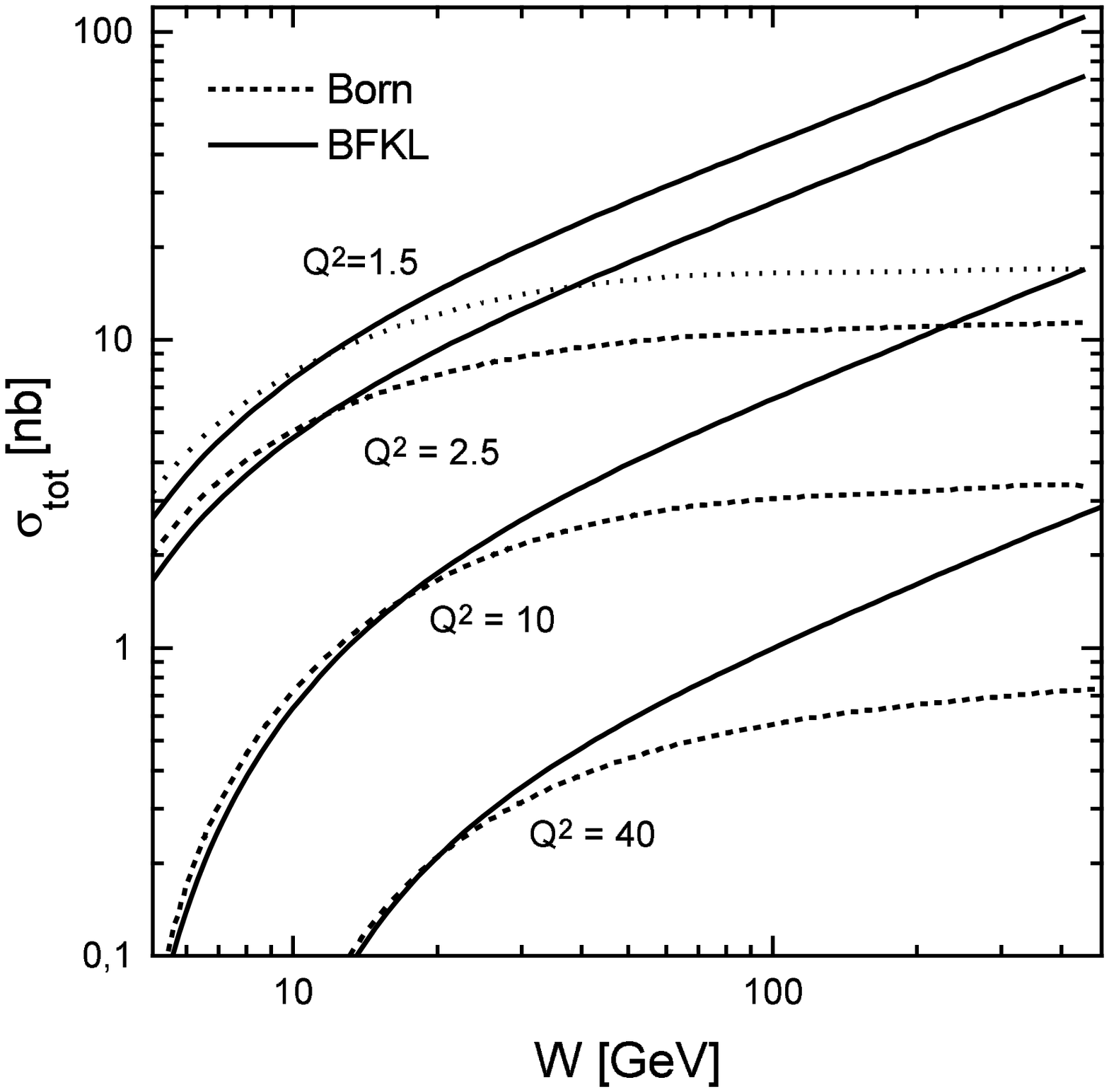}} \\
\end{center}
\caption{ \small Energy dependence of the cross-section
$\sigma^{TT} _{\gamma^* \gamma^*}(Q_1 ^2, Q_2 ^2, W^2)$
for the process $\gamma^*(Q_1 ^2) \gamma^*(Q_2^2)  \to hadrons$
for various choices of photon virtualitities $Q^2 = Q_1 ^2 = Q_2 ^2$:
a)~Complete i.e. including soft and QCD pomeron contributions  results 
corresponding to Eq.~(\ref{csx}). 
For each choice of the virtualitity $Q^2$  
four curves are shown taking into account hard effects only (``hard part''),
hard amplitude with soft pomeron contributions added in the source term
of the BFKL equation (``mixed''),
the full cross-section including both soft and hard pomeron
contributions (``full result''). We also show the ``full result'' with the
low scale of $\alpha_s$ in the impact factors:
$\mu^2 = (k^2 + m_q^2)/4$.
b)~Comparison of the complete contribution of the perturbative QCD pomeron  
to the cross-section $\sigma^{TT} _{\gamma^* \gamma^*}(Q ^2, Q ^2, W^2)$
(continous line) with its Born term component corresponding to the two gluon
exchange mechanism (dotted line).  }
\end{figure}

In Fig.~2a we show our results for 
$\sigma^{TT}_{\gamma^* \gamma^*}(Q_1 ^2, Q_2 ^2, W^2)$ plotted
as the function of the CM energy $W$ for three different values of  $Q^2$
where $Q_1^2 = Q_2^2 =Q^2$.  We plot in this figure:
\begin{enumerate}
\item the pure perturbative QCD (i.e. ``hard'') contribution obtained 
from solving the BFKL
equation with the consistency constraint included (see Eq.~(\ref{bfklcc}))
and with the inhomogeneous term containing only the QCD impact factor defined
by equations (\ref{phii0},\ref{tgt0},\ref{tgl0});
\item the ``mixed'' contribution generated by the BFKL equation (\ref{bfklcc})
with the soft pomeron contribution defined by equations (\ref{stophi},
\ref{inhos}) included in the inhomogeneous term;
\item
The ``full" contribution which also contains the soft pomeron term (\ref{sp}).
\end{enumerate}
We also show results obtained by changing the scale of the strong coupling
$\alpha_s$  in the impact factors from $k^2+m_q^2$ to $(k^2+m_q^2)/4$.
The scale of $\alpha_s$ in the BFKL equation was set equal to $k^2$ 
in both the cases.\\

We see from this figure that the effects of the soft pomeron contribution
are non-negligible at low and moderately large values of $Q^2 < 10\gev^2$
and for moderately large values of $W < 100 \gev$.
The QCD pomeron however dominates already at $Q^2=40 \gev^2$.
We also see from this figure that for low energies $W<40 \gev$ the
phase-space limitations (cf. Eqs.~(\ref{kmax}) and (\ref{ximin})) 
are very important. For $W>40 \gev$ or so one observes that
the cross-section exhibits the effective power-law behaviour
$\sigma^{TT} _{\gamma^* \gamma^*}(Q^2, Q^2, W^2) \sim (W^2)^{\lambda}$.
The (effective) exponent increases weakly
with increasing $Q^2$ and varies from $\lambda=0.28$ for $Q^2=2.5\gev^2$ to
$\lambda=0.33$ for $Q^2 = 40 \gev^2$.  This (weak) dependence of the
effective exponent $\lambda$ on  $Q^2$ is the result of the interplay 
between soft and hard pomeron contributions,
where the former becomes less important at large $Q^2$.
The shape of the remaining cross sections $\sigma_{\gamma^* \gamma^*}^{TL}$, 
$\sigma_{\gamma^* \gamma^*}^{LT}$ and $\sigma_{\gamma^* \gamma^*}^{LL}$ as 
functions of $W$ is the same as that of $\sigma_{\gamma^* \gamma^*}^{TT}$.  
They do however differ in their relative normalisation.\\

In Fig.~2b we compare the QCD pomeron contribution 
to the cross-section 
$\sigma^{TT}_{\gamma^* \gamma^*}$
with the Born term,  which corresponds 
to the two-gluon exchange mechanism.  The latter contribution 
is given by equation (\ref{csx}) with the functions $\Phi_i(k^2,Q_1^2,x\xi)$
 approximated 
by the impact factors $\Phi_i^0(k^2,Q_1^2,x\xi)$ defined by equations 
(\ref{phii0}), (\ref{tgt0}) and (\ref{tgl0}).  
For large values of the CM energy $W$ the two-gluon exchange 
mechanism gives energy independent contribution to the cross-section 
$\sigma^{TT}_{\gamma^* \gamma^*}$. In the low energy region  
its energy dependence 
(i.e. the onset of the constant asymptotic behaviour) is controlled 
by phase space effects embodied in equations (\ref{csx}), (\ref{kmax}) and 
(\ref{ximin}). 
We can also see from this figure that the dominance of the BFKL pomeron over 
its Born term is  delayed to higher energies.   In the low energy region 
the cross-section  corresponding to the solution of the modiffied BFKL 
equation (\ref{bfklcc}) is even smaller than 
that which is given by the Born term.  This is caused by the fact,  which we 
have already mentioned above,   that for moderately small  
values of $x_g/x^{\prime}$ the consistency constraint (\ref{cc}) supresses 
the (positive) real emission contribution to the BFKL kernel 
while leaving unaffected  the (negative)  virtual corection term 
(cf. Eq.~(\ref{bfklcc})).      
\\

\noindent
\begin{table}[h]

\begin{center}
\begin{tabular}{||c|c|ccc|cc||}
\hline\hline
 &  \multicolumn{6}{|c||}{$\langle d\sigma / dY \rangle$ [fb] }\\
\cline{2-7}
  & & \multicolumn{5}{c||}{ Theory (BFKL+SP)} \\
\cline{3-7}
$\Delta Y$ & Data --- QPM & \multicolumn{3}{c|}{$\alpha_s[(k^2+m_q^2)/4]$} &
       \multicolumn{2}{c||}{$\alpha_s (k^2+m_q^2)$} \\
&    & Born & Hard & Hard + SP & Hard & Hard + SP \\
\hline
\multicolumn{7}{||c||}{91 GeV} \\ \hline
2 -- 3 & $480 \pm 140 \pm 110$  & 91 & 76 & 206 & 34 & 163\\
3 -- 4 & $240 \pm 60 \pm 50  $  &125 & 114& 237 & 53 & 173\\
4 -- 6 & $110 \pm 30 \pm 10  $  & 56 & 60 & 109 & 29 & 74\\ \hline
\multicolumn{7}{||c||}{183 GeV} \\ \hline
2 -- 3 & $180 \pm 120 \pm 50$   & 55 &  51 & 68  & 25 & 42\\
3 -- 4 & $160 \pm 50 \pm 30  $  & 67 &  70 & 86  & 34 & 49\\
4 -- 6 & $120 \pm 40 \pm 20  $  & 53 &  70 & 85  & 35 & 47\\ \hline\hline
\end{tabular}
\end{center}

\caption{\small Comparison of the theoretical results to L3 data for
$e^+ e^- \to e^+ e^- + hadrons$
with  $E_{tag} > 30$ GeV, 30 mrad $ < \theta_{tag}  <$ 66 mrad. We show in
the table $d\sigma / dY$ binned in $Y$ obtained from experiment and
the results of our calculation which take into account two-gluon exchange
(Born approximation) perturbative pomeron (hard) and both perturbative and
soft pomerons (hard + SP) for two different choices of scale of 
the $\alpha_s$ in impact factors and 
for $e^+e^-$ CM energy 91~GeV and 183~GeV.}

\end{table}

We have also calculated the total cross-section of the process
$e^+ e^- \rightarrow e^+ e^- + hadrons$ (see Eq.~(\ref{conv})) for LEP1
and LEP2 energies and confronted results of our calculation with the
recent experimental data obtained by the L3 collaboration at LEP \cite{L3}.
Comparison of our results with experimental data is summarised in Tab.~1.
We show comparison for  $d\sigma/dY$, where $Y=\ln(W^2/Q_1Q_2)$
with subtracted Quark Parton Model (QPM)  contribution.
We see that the contamination of the cross-section by soft pomeron
is substantial.  The data do also favour the smaller value of the scale of
$\alpha_s$.   We also show in this table theoretical predictions in which the
QCD pomeron was approximated by the two-gluon exchange contribution.  
%
%
We notice that throughout  (almost) entire $Y$ range, which is being probed at 
LEP the QCD pomeron is dominated by its two-gluon exchange part and that the 
effects of the BFKL enhancement are not visible.  
This is caused by the fact  that  in the region $Y < 4$  the two gluon 
exchange dominates 
the cross-sections $\sigma^{ij}_{\gamma^* \gamma^*}$ (see Fig.~2b) and that 
the BFKL enhancement is delayed to higher values of $Y$.  We can notice  
onset of this enhancement in the last bin $4<\Delta Y<6$ by it 
is still a very weak effect.  We conclude that the energies which are 
accessible at LEP are insufficient for probing the BFKL enhancement.  
This is due to the fact that the subleading 
effects  delay the onset of the power-law BFKL behaviour of the 
cross-sections $\sigma^{ij} _{\gamma^* \gamma^*}$ to higher 
energies than those which are probed at LEP. 
\\

\noindent
\begin{table}[hbpt]

\begin{center}
\begin{tabular}{||c|ccc|c||}
\hline\hline
$\theta_{min}$ --- $\theta_{max}$ &
\multicolumn{3}{c|}{$\sigma (e^+e^- \to e^+ e^-  + hadrons)$ [fb]} &
Events / year \\
  &Born &  Hard & Full (LS)& Full (LS) \\ \hline
10--20 & 134 & 365 & 450 & 9000 \\
20--30 & 16  & 41  & 46  & 920  \\
30--40 & 3.5 & 8   &  9  & 180  \\
40--50 & 1.1 & 2.3 & 2.5 & 50   \\
50--70 & 0.6 & 1.1 & 1.3 & 26  \\ \hline
30--70 & 5.2 & 11  & 13  & 260 \\  \hline\hline
\end{tabular}
\end{center}

\caption{\small Predictions for TESLA at $e^+e^-$ energy equal to 500~GeV.
Cross-sections for $e^+ e^- \to e^+ e^-  + hadrons$ with tagged electrons
$E_{tag} > 30 {\rm GeV}$, $y_i >0.1$, 2.5 GeV$^2 < Q_i ^2  < 300$~GeV$^2$,
$2<\ln [W^2 / (Q_1 Q_2)] < 10$, $\theta_{min} < \theta_{tag} < \theta_{max}$.
Results of the calculation with the low scale of $\alpha_s$
in impact factors: two-gluon exchange (Born approximation), hard and full
(hard+soft) contributions and
the expected number of events per year, assuming
the integrated luminosity per year to be ${\cal L} = 20{\rm fb}^{-1}$.
}

\end{table}

In Tab.~2 we show results of our estimate for the cross-section for the
process $e^+ e^- \rightarrow e^+ e^- + hadrons$ with tagged $e^+ e^-$ in 
the final state for the total CM energy of the $e^+ e^-$ system equal to 
500~GeV.  We can see that in this very high energy
region the QCD pomeron dominates
over the soft pomeron contribution even for very low tagging angles.
These cross-sections are also  bigger by about a factor equal to 2--3
than the two-gluon exchange ``background".\\

Different configurations of the  virtual photon polarizations 
contribute to the QCD pomeron part of the cross-section for the process 
$e^+e^- \rightarrow e^+e^- + hadrons$ with the following approximate relative 
normalizations: $(TT) : (TL+LT) : (LL) = 9:6:1$.  This means that the 
transversely polarized virtual photons alone give about 60\% 
of the QCD pomeron part of the cross-section.   
\\

To sum up we have estimated the total $\gamma^* \gamma^*$ cross-section
and its impact on the cross-section of the process $e^+  e^- \rightarrow
e^+  e^- + hadrons$ with  tagged $e^+ e^-$ in the final state for LEP energies
and for the energy range which may become accesible in TESLA and in other
future $e^+e^-$ linear colliders.
We based our calculations on the QCD pomeron exchange generated by the 
BFKL equation (\ref{bfklcc}) with the subleading effects which follow from the 
consistency constraint (\ref{cc}).  
We have also included  the ``soft" pomeron term which gives
important contribution at moderately large values of $Q^2$ and $W^2$.
The subleading effects in the BFKL equation significantly 
reduce the QCD pomeron
intercept and the magnitude of the corresponding cross-sections.   The total
$\gamma^* \gamma^*$  cross-section and the cross-section of the process
$e^+  e^- \to e^+  e^- + hadrons$ with  tagged $e^+ e^-$ in the final state are
smaller than those corresponding to the LO BFKL equation. For sufficiently
high energies however they are significantly bigger than the cross-sections 
which would follow from the two-gluon exchange mechanism.  
We confronted our theoretical predictions with recent
experimental results from LEP  obtaining fairly reasonable agreement 
with the data. 
We find that the BFKL effects should become clearly visible 
for energies which can become accessible in future linear colliders.

\section*{Acknowledgments}
 We thank Albert De Roeck for his interest in this work and useful
discussions. This research was partially supported
by the Polish State Committee for Scientific Research (KBN) grants
2~P03B~184~10, 2~P03B~89~13, 2~P03B~084~14 and by the
EU Fourth Framework Programme `Training and Mobility of Researchers', Network
`Quantum Chromodynamics and the Deep Structure of Elementary Particles',
contract FMRX--CT98--0194.


\begin{thebibliography}{9999}
\bibitem{BRODSKY}S.J. Brodsky, F. Hautmann, D.A. Soper, Phys. Rev. {\bf D56} (1997) 6957;
Phys. Rev. Lett. {\bf 78} (1997) 803 (Erratum-ibid. {\bf 79} (1997) 3544).
%
\bibitem{BARTELSDR}J. Bartels, A. De Roeck, H. Lotter, Phys. Lett.
{\bf B389} (1996) 742;   J. Bartels , A. De Roeck, C. Ewerz,
H. Lotter, hep-ph/9710500.
\bibitem{CZYZFL}A. Bia\l{}as, W. Czy\.z, W. Florkowski, Eur. Phys. J. {\bf C2} (1998) 683;
W. Florkowski, Acta Phys. Polon. {\bf 28} (1997) 2673;
A.~Donnachie, H.G.~Dosch, M. Rueter, Phys.~Rev.~{\bf~D59} (1999) 74011;
M.~Boonekamp et al. hep-ph/9812523.
%
\bibitem{BFKL}E.A. Kuraev, L.N.Lipatov and V.S. Fadin, Zh. Eksp. Teor. Fiz.
{\bf 72} (1977) 373 (Sov. Phys. JETP {\bf 45} (1977) 199);
Ya. Ya. Balitzkij and L.N. Lipatov, Yad. Fiz. {\bf 28} (1978) 1597 (Sov. J.
Nucl. Phys. {\bf 28} (1978) 822);
J.B. Bronzan and R.L. Sugar, Phys. Rev. {\bf D17} (1978) 585;
T. Jaroszewicz, Acta. Phys. Polon. {\bf B11}
(1980) 965; L.N. Lipatov, in "Perturbative QCD", edited
by A.H. Mueller, (World Scientific, Singapore, 1989), p. 441.
%
\bibitem{GLR} L.N. Gribov, E.M. Levin and M.G. Ryskin,
Phys. Rep. {\bf 100} (1983) 1.
%
\bibitem{BFKLNL} M. Ciafaloni, G. Camici, Phys. Lett. {\bf B386} (1996) 341;
ibid. {\bf B412} (1997) 396; Erratum -- ibid. {\bf B417}
(1998) 390; hep-ph/9803389; M. Ciafaloni, hep-ph/9709390;
V.S. Fadin, M.I. Kotskii, R. Fiore, Phys. Lett. {\bf B359} (1995) 181;
V.S. Fadin, M.I. Kotskii, L.N. Lipatov, hep-ph/9704267; V.S. Fadin, R. Fiore,
A. Flachi, M. Kotskii, Phys. Lett. {\bf B422} (1998) 287;
V.S. Fadin, L.N. Lipatov, hep-ph/9802290.
%
\bibitem{RESUM} D.A.~Ross, Phys. Lett. {\bf B431} (1998) 161;
G.P.~Salam, JHEP {\bf 9807} (1998), 19; hep-ph/9806482;
M.~Ciafaloni, D.~Colferai hep-ph/9812366; S.J.~Brodsky et al.
hep-ph/9901229; C.R.~Schmidt hep-ph/9904368.
%
\bibitem{KMSG} B. Andersson, G. Gustafson, H. Kharraziha, J. Samuelsson,
Z. Phys. {\bf C71} (1996) 613;   J. Kwieci\'nski, A.D. Martin, P.J. Sutton,
Z. Phys. {\bf C71} (1996) 585.
%
\bibitem{KMS} J.Kwieci\'nski, A.D. Martin and  A.M. Sta\'sto,
Phys. Rev. {\bf D56} (1997) 3991.)
\bibitem{DLTOTCX} A. Donnachie and P.V. Landshoff, Phys. Lett. {\bf B296} (1992) 227.
\bibitem{DLDIS} A. Donnachie and P.V. Landshoff, Z. Phys. {\bf C61}
(1994) 139.
%
\bibitem{L3} L3 collaboration, (M. Acciari et al) CERN-EP-98-205 (1998).
%
\end{thebibliography}
\end{document}